\documentclass[fleqn,usenatbib]{mnras}

\usepackage{newtxtext,newtxmath}

\usepackage[T1]{fontenc}
\usepackage{ae,aecompl}
\usepackage{amsmath}

\usepackage{graphicx}	
\usepackage{amsmath}	
\usepackage{amssymb}	
\usepackage{multirow}
\usepackage{rotating}
\usepackage{dblfloatfix}
\usepackage{tabularx}
\usepackage{placeins}
\usepackage{pdflscape}

\usepackage[para,online,flushleft]{threeparttable}

\usepackage{hyperref}
\usepackage{lscape}

\hypersetup{
  colorlinks=true,
  linkcolor=blue,
  filecolor=blue,      
  urlcolor=blue,
  citecolor=blue,
}

\title[Tracing circumnuclear dense gas in H\textsubscript{2}O maser galaxies]{Tracing circumnuclear dense gas in H\textsubscript{2}O maser galaxies}

\author[Farhan, Ercan \& F. Tombesi]{ 
A. Farhan$^{1}$\thanks{{\color {blue} E-mail: ahlamfarhan@yahoo.com (AF)}}, E. N. Ercan$^{1}$\thanks{{\color {blue} ercan@boun.edu.tr (ENE)}}
and F. Tombesi$^{2,3,4,5}$\thanks{{\color {blue} francesco.tombesi@roma2.infn.it (FT)}}
\\
$^{1}$Department of Physics, Bogazici University, Istanbul, 34342, Turkey\\
$^{2}$Department of Physics, Tor Vergata University of Rome, Via della Ricerca
Scientifica 1, 00133 Rome, Italy
\\$^{3}$INAF - Astronomical Observatory of Rome, Via Frascati 33, 00044 Monte
Porzio Catone, Rome, Italy
\\$^{4}$Department of Astronomy, University of Maryland, College Park, MD 20742,
USA\\$^{5}$X$-$ray Astrophysics Laboratory, NASA/Goddard Space Flight Center,
Greenbelt, MD 20771, USA
}

\date{Accepted XXX. Received YYY; in original form ZZZ}

\pubyear{2020}

\begin{document}
\label{firstpage}
\pagerange{\pageref{firstpage}--\pageref{lastpage}}
\maketitle


\begin{abstract}
A sample of 30 H$\textsubscript{2}$O extra-galactic maser galaxies with their published HCN(J=1$-$0) and HCO+(J=1$-$0) observations has been compiled to investigate the dense gas correlation with H\textsubscript{2}O maser emission. Our sample number exceeds the size of the previous HCN samples studied so far by a factor of three, and it is the first study on the possible relation with the dense gas tracer HCO+. We find a strong correlation between normalized H\textsubscript{2}O maser emission luminosity (L\textsubscript{H\textsubscript{2}O}/L\textsubscript{CO}) and normalized HCO+ line luminosity (L\textsubscript{HCO+}/L\textsubscript{CO}). Moreover, a weak correlation has been found between L\textsubscript{H\textsubscript{2}O}/L\textsubscript{CO} and normalized HCN line luminosity (L\textsubscript{HCN}/L\textsubscript{CO}). The sample is also studied after excluding Luminous and Ultraluminous infrared galaxy (U)LIRG sources, and the mentioned correlations are noticeably stronger. We show that 'Dense gas' fractions as obtained from HCN and HCO+ molecules tightly correlate with maser emission, especially for galaxies with normal IR luminosity(L\textsubscript{IR}< 10\textsuperscript{11}L\textsubscript{\(\odot\)}) and we show that HCO+ is a better 'dense gas' tracer than HCN. Further systematic studies of these dense gas tracers with higher transition level lines are vital to probe megamaser physical conditions and to accurately determining how maser emission interrelates with the dense gas.  

\end{abstract}



\section{Introduction}
Since the discovery of extragalactic water maser galaxies (emitting at $\nu$  = 22.2~GHz, $\lambda$ = 1.35~cm) by \cite{1977A&A....54..969C}, several surveys have been conducted using different constraints on the chosen samples, e. g. X-ray spectrum, hydrogen column density, and nuclear radio continuum. Such constraints have been considered to increase the detection rate.

Megamaser host galaxies have been found to have complex X-ray spectra together with higher X-ray luminosities than non-maser galaxies (e.g. \citealt{Kondratko-2006}; \citealt{Leiter2017}). Also the megamaser sources have been found to preferably exist in galaxies with high hydrogen column density (N\textsubscript{H} $\geq$10\textsuperscript{22}~cm\textsuperscript{-2}) due to the connection of masing spots with a highly obscured region near the accretion disk as discussed by \cite{Braatz1997}, \cite{Zhang2006}, \cite{Greenhill:2008ra}, and \cite{Castangia:2013nca}. Most commonly, maser emission is observed in Seyfert 2  or Low Ionization Nuclear Emission-Line Region (LINER ) type of galaxies (see e.g. \citealt{Braatz1997}; \citealt{Lo2005}). Recent studies show that megamaser galaxies have 2-3 orders of magnitude higher nuclear radio continuum luminosities than non-masers (see e.g. \citealt{Zhang2012}; \citealt{Liu2016}).

However, the detection  rate found so far  is still low: of the order of 3 per cent. In fact, from roughly 6000 galaxies surveyed for H\textsubscript{2}O megamasers (H\textsubscript{2}OMM) there are only 180 detected  as recently reported by \cite{Kuo-2020}.

The low detection rate of H\textsubscript{2}OMM was the motivation for several researchers to suggest other possible search criteria (\citealt{Zhang2006}; \citealt{Greenhill:2008ra}; \citealt{Zhang-2009}; \citealt{Ramolla-2011}; \citealt{Zhu-2011}; \citealt{Constantin-2012}; \citealt{Van den Bosch-2016}; \citealt{Liu-2017}; \citealt{Kuo2018}). \\
 It was proposed by \cite{Claussen1986} that luminous H\textsubscript{2}OMM form in dense gas clouds (n(H\textsubscript{2})$\geq$10\textsuperscript{4}~cm\textsuperscript{-3}) in the circumnuclear disks (CND) of Active Galactic Nuclei (AGN), fuelled by accretion.
  Therefore, dense gas tracers were suggested to be a suitable indicator of H\textsubscript{2}O masers (\citealt{Zhang-2009}) since they track high density and temperature molecular clouds excited by collisions. Amongst the most abundant dense gas tracers in galaxies is HCN with a large dipole moment ($\mu \approx$ 2.98 Debye D) and  it can be thermalised by collisions at high densities, with a critical density of n\textsubscript{crit} $\sim$ 10\textsuperscript{6}~cm\textsuperscript{-3} at which the spontaneous emission rate equals  to the collisional transition rate (see e.g. \citealt{Baan2008}).  HCO+ is another dense gas tracer with a dipole moment $\mu \sim$ 3.92  D, and   n\textsubscript{crit}$\sim$ 10\textsuperscript{5}~cm\textsuperscript{-3}.
 
 Thanks to the large HCN survey published by \cite{Gao2004}, \cite{HUANG} were able to study the possible correlation between dense gas tracers and  maser emission, finding a positive correlation. However, their sample was limited to 13 galaxies. In this work, we  increased the sample to 30 galaxies with HCN observations, moreover, this is the first work studying the possible correlations between maser emission and another dense gas tracer, specifically HCO+.
 
In Section~2 we describe the sample and its corresponding data. Analyses and results of non-parametric statistical approaches to the sample are presented in Section~3. Discussion of the results is presented in Section~4 while main results are summarized in Section~5. Throughout our analyses we adopt flat cosmology parameters  of ($\Omega$ \textsubscript{m} = 0.27, $\Omega$ \textsubscript{$\Lambda$} = 0.73, and H\textsubscript{0}= 71~kms$^{-1}$Mpc$^{-1}$); \cite{Spergel2007}. Wherever CO, HCN, or HCO+ appear in this work, we mean the first rotational excitation J-transition emission lines, in other words, CO(J=1-0), HCN(J=1-0) and HCO+(J=1-0), respectively, unless other line is explicitly mentioned.   

\section{Dataset}
There are about 180 known H\textsubscript{2}O galaxies so far. A catalog of 22 GHz H\textsubscript{2}OMM galaxies was updated by Jim Braatz at the National Radio Astronomy Observatory (NRAO)\footnote{https://safe.nrao.edu/wiki/bin/view/Main/PublicWaterMaserList}. We checked the line emissions of HCN(J=1-0), HCO+(J=1-0), and \textsuperscript{12}CO(J=1-0) available in the literature for these galaxies, and compiled our  sample in Table \textsc{1}.

 Most of the HCN data were taken from \cite{Liu2015}. For the unreported values, we estimated the HCN and CO luminosities using the same equation as in \cite{Liu2015}, as reported in our eq.(1). The references for all sources are provided in Table \textsc{1}. 
\cite{Liu2015} used the following equation to calculate the line luminosity for CO and HCN:
\begin{equation}
L\textsubscript{HCN,CO} (\textbf{K}kms\textsuperscript{-1}pc\textsuperscript{2})\textsuperscript{-1} = 3.25 \times10^{7} \times S\textsubscript{HCN,CO} \Delta \nu \times \nu \textsubscript{obs} \textsuperscript{-2} \times D\textsubscript{L}^2,
\end{equation}
where S\textsubscript{HCN,CO} $\Delta \nu$  is the velocity integrated flux of HCN and/or CO lines in  Jykms$^{-1}$, $\nu$\textsubscript{obs} is the observed frequency for HCN (88.63 GHz), for CO (115.27 GHz), and D\textsubscript{L} is the luminosity distance in Mpc. HCO+ luminosities were also calculated using eq.(1) with $\nu$\textsubscript{obs} = 89.19~GHz. The uncommon unit of L\textsubscript{HCN,CO,HCO+} (Kelvin kilometer parsec squared per second) is due to spectral observation methods used in millimeter range, where the integration of brightness temperature with respect to the emission velocity is usually used to express line intensity \cite{Solomon-1997}. 
$\log_{10}$(L\textsubscript{HCN}/\textbf{K}kms\textsuperscript{-1}pc\textsuperscript{2}) ranges from 5.6 to 9.1 with a mean of 7.5 , while $\log_{10}$(L\textsubscript{HCO+}/\textbf{K}km s\textsuperscript{-1}pc\textsuperscript{2}) ranges from 5.9 to 9 with a mean of 7.4. Our maser sample contains 17 megamaser galaxies (MM, $\log_{10}$(L\textsubscript{H\textsubscript{2}O}/\(\textup{L}_\odot\)) $\geq$ 1) and 13 kilomaser galaxies (KM, $\log_{10}$(L\textsubscript{H\textsubscript{2}O}/\(\textup{L}_\odot\)) $<$1 ), the maximum value of $\log_{10}$(L\textsubscript{H\textsubscript{2}O}/\(\textup{L}_\odot\)) is 3.3 for the centre of the ultra luminous infrared galaxy (ULIRG) UGC5101, and the minimum is -2 in the nearby starburst galaxy IC342, the average of $\log_{10}$(L\textsubscript{H\textsubscript{2}O}/\(\textup{L}_\odot\)) is 1.

\section{Analysis of the dataset}
We use non-parametric statistics in order to analyse our relatively small sample. Most often, non-parametric statistics are based on ranks of the two variables (let's say x and y) rather than their values, where x\textsubscript{i} and y\textsubscript{i} are ordered increasingly and given ranks, then an analysis is carried out on these ranks rather than the data themselves. The power of using rank order analyses is shown when we have outliers (or anomalous points), avoiding the use of means of the variables x\textsubscript{i} and y\textsubscript{i} and their standard deviations (which is essentially the case in parametric analyses). This allows us to  include outliers in the analysis without affecting the actual regression line of the relation (\citealt{Kandalyana}; \citealt{Kandalyanb}).
 The Spearman's rank correlation coefficient, $\rho$, is the parametric Pearson correlation coefficient for the ranking of the x\textsubscript{i} and y\textsubscript{i} variables (see e.g. \citealt{Obremski1981}; \citealt{Myers}) and can be calculated as follows:

\begin{equation} 
\rho = \frac{ \sum_{i=1}^{n} (R_{x_i}-\overline{R_x})(R_{y_i}-\overline{R_y})}{\sqrt{\sum_{i=1}^{n}(R_{x_i}-\overline{R_x})^2.\sum_{i=1}^{n}(R_{y_i}-\overline{R_y})^2}},
\end{equation}

where the ranks R\textsubscript{x\textsubscript{i}} and R\textsubscript{y\textsubscript{i}} are the position of the i\textsuperscript{th} x and i\textsuperscript{th} y variables in the sample, respectively, after ordering it, n is the sample size, $\overline{R\textsubscript{x}}$ and $\overline{R\textsubscript{y}}$ are the mean values of the ranks of the variables x and y, respectively.
We calculated the Spearman rank-order correlation coefficients between the luminosity fractions given in Table \textsc{1} and present the non-parametric statistical results in Table \textsc{2}.

 Table \textsc{2} shows the  Spearman's rank correlation coefficient ,$\rho$, between the variables and P-value of the corresponding $\rho$. Also shown in Table \textsc{2} is the result of Kendall-Theil Line regression (\citealt{Kandalyana}) including the slope of the linear regression line S, intercept I, 95\% confidence interval for the slope, and the median deviation d.
 
It should be taken into  consideration that luminosities of maser and dense gas tracers depend significantly on the distance to the sources. Thus, if both of the variables being analysed depend on the redshift z, the so-called Malmquist effect (\citealt{Butkevich2005}) should be eliminated first. For this purpose, we normalised the luminosities to L\textsubscript{CO} to get the intrinsic relation between the variables without being affected by the distance. Using L\textsubscript{CO(1-0)} as a normaliser is very helpful because this molecule has a low dipole moment $\mu \approx$ 0.11 D and low critical density, n\textsubscript{crit} 	$\approx$ 2.2x10\textsuperscript{3}~cm\textsuperscript{-3} at T$\approx$30K as it is estimated by \cite{Yang2010}. These conditions are essentially satisfied in any molecular gas cloud, that is why CO(1-0) is used as a lower-density gas tracer. Normalising dense gas tracer luminosity to L\textsubscript{CO} gives us the dense gas ratio in galaxies.
The Central R Archive Network {\sc r cran}'s package {\sc ggpubr}\footnote{https://cran.r-project.org/web/packages/ggpubr/index.html} was used to calculate Spearman correlations \cite{Kassambara2020}, and {\sc KTRLine} software\footnote{https://pubs.usgs.gov/tm/2006/tm4a7/} \cite{Granato2006} for Kendall-Theil analyses.
 
  \newpage
 \onecolumn
 \begin{landscape}
\begin{table}
\centering

\caption{The 30 H\textsubscript{2}OMM data sample}.
\label{tab:my-table}
\resizebox{1.2\textwidth}{!}{
\begin{tabular}{lllllllllll}

\hline
\\
Galaxy name & z&D\textsubscript{L} & $ \log_{10}$ (L$_{\mathrm{H\textsubscript{2}O}} / L_{\sun}$)  & 
$ \log_{10} (L_{\mathrm{HCN}} / \textbf{K}kms^{-1}pc\textsuperscript{2}) $ & $ \log_{10} (L_{\mathrm{HCO+}} / \textbf{K}kms\textsuperscript{-1}pc\textsuperscript{2}) $ & $ \log_{10} (L_{\mathrm{CO}} / \textbf{K}kms\textsuperscript{-1}pc\textsuperscript{2}) $ & $\log_{10}$(L\textsubscript{H\textsubscript{2}O}/L\textsubscript{CO}) & $\log_{10}$(L\textsubscript{HCN}/L\textsubscript{CO}) & $\log_{10}$(L\textsubscript{HCO+}/L\textsubscript{CO}) &References \\
\\
\hline\hline
NGC 23                & 0.015231      & 62.4   & 2.26  & 7.81 & **   & 9.21 & -6.95  & -1.4  & **    & 1,3,**,3   \\
NGC 17 (NGC34)        & 0.019617      & 80.92  & 0.95  & 8.95 & **   & 9.55 & -8.6   & -0.6  & **    & 2,3,**,3   \\
NGC 253               & 0.000811      & 4.11   & -0.7  & 7.82 & 7.37 & 9.18 & -9.88  & -1.36 & -1.82 & 1,3,4,3    \\
NGC 520               & 0.007609      & 30.59  & 0     & 7.79 & **   & 9.46 & -9.46  & -1.67 & **    & 1,3,**,3   \\
NGC 613               & 0.00494       & 20.9   & 1.3   & 7.43 & **   & 8.54 & -7.24  & -1.11 & **    & 1,5,**,11  \\
NGC 1068              & 0.003793      & 15.46  & 2.2   & 8.48 & 8.15 & 9.47 & -7.27  & -0.99 & -1.32 & 1,3,6,3    \\
IC 342                & 0.000103      & 4.32   & -2    & 7.6  & 6.54 & 9.21 & -11.21 & -1.61 & -2.67 & 1,3,6,3    \\
NGC 2146              & 0.002979      & 15.34  & 0     & 8.01 & 7.28 & 9.16 & -9.16  & -1.15 & -1.88 & 1,3,6,3    \\
NGC 2273              & 0.006138      & 28.54  & 1.51  & 7.32 & 6.84 & 7.88 & -6.37  & -0.56 & -1.03 & 1,3,7,3    \\
He 2-10               & 0.002912      & 12     & -0.15 & 6.6  & 6.81 & 8.32 & -8.47  & -1.72 & -1.51 & 1,3,8,3    \\
UGC 5101              & 0.039367      & 164.93 & 3.28  & 9.07 & 8.48 & 9.69 & -6.41  & -0.62 & -1.21 & 1,3,9,3    \\
M 82 (NGC3034)        & 0.000677      & 4.89   & 0     & 7.65 & 7.26 & 9.01 & -9.01  & -1.38 & -1.77 & 1,3,6,3    \\
NGC 3079              & 0.003723      & 19.51  & 2.7   & 7.59 & 7.63 & 9.01 & -6.31  & -1.42 & -1.38 & 1,12,12,11 \\
NGC 3256              & 0.009354      & 35.86  & 1     & 8.39 & 8.5  & 9.67 & -8.67  & -1.28 & -1.17 & 1,3,6,3    \\
NGC 3556              & 0.002332      & 12.09  & 0     & 6.28 & 6.48 & 8.57 & -8.57  & -2.29 & -2.09 & 1,12,12,3  \\
NGC 3620              & 0.005604      & 19.18  & 0.6   & 7.82 & 6.88 & 8.71 & -8.11  & -0.89 & -1.83 & 1,3,13,3   \\
Arp 299               & 0.010411      & 48.02  & 2.1   & 8.42 & 8.26 & 9.54 & -7.44  & -1.12 & -1.28 & 1,3,6,3    \\
NGC 4038 (NGC4039)    & 0.005688      & 21.1   & 0.9   & 7.2  & 7.68 & 9.1  & -8.2   & -1.9  & -1.41 & 1,3,6,3    \\
NGC 4051              & 0.002336      & 9.9    & 0.3   & 5.57 & 5.92 & 7.63 & -7.33  & -2.06 & -1.71 & 1,7,7,11   \\
NGC 4258              & 0.001494      & 6.1    & 1.9   & 6.39 & 6.61 & 7.76 & -5.86  & -1.37 & -1.15 & 1,14,14,15 \\
NGC 4388              & 0.008419      & 34.8   & 1.08  & 6.93 & 7.29 & 8.68 & -7.6   & -1.75 & -1.39 & 1,14,14,11 \\
NGC 4945              & 0.001878      & 5.2    & 1.7   & 7.67 & 7.7  & 9.2  & -7.5   & -1.53 & -1.5  & 1,3,6,3    \\
NGC 5128(Centaurus-A) & 0.001823 & 7.7    & 0     & 6.26 & 6.29 & 8.2  & -8.2   & -1.94 & -1.91 & 2,16,16,11 \\
M 51 (NGC5194)        & 0.001544      & 7.53   & -0.22 & 6.78 & **   & 9.15 & -9.37  & -2.37 & **    & 1,3,**,3   \\
NGC 5256              & 0.027863      & 121.77 & 1.51  & 8.11 & 8.42 & 9.88 & -8.37  & -1.77 & -1.46 & 1,3,17,3   \\
NGC 5347              & 0.007789      & 32.18  & 1.51  & 7.2  & **   & 7.59 & -6.08  & -0.39 & **    & 1,3,**,3   \\
Circinus              & 0.001448      & 6.1    & 1.3   & 7.23 & **   & 8.41 & -7.11  & -1.18 & **    & 1,19,**,11 \\
NGC 5506              & 0.006181      & 25.5   & 1.7   & 6.63 & 6.93 & 7.8  & -6.1   & -1.17 & -0.9  & 1,14,14,20 \\
NGC 6240              & 0.02448       & 106.71 & 1.6   & 8.89 & 8.99 & 9.92 & -8.32  & -1.03 & -0.93 & 1,12,21,3  \\
NGC 7479              & 0.007942      & 33.56  & 1.28  & 7.97 & **   & 9.36 & -8.08  & -1.39 & **    & 1,3,**,3  \\
\hline
\end{tabular}}
\begin{tabular}{p{22cm}}
\begin{tablenotes}
\item {\bf Note.}
Columns  shown in the Table are: (1) Name of the galaxy. (2) Redshift, z taken from Preferred redshift from NASA/IPAC Extragalactic Database (NED)(https://ned.ipac.caltech.edu/). (3) Luminosity distance, D\textsubscript{L} in Mpc, when available, taken from \cite{Liu2015}, otherwise calculated using NED cosmology calculator (http://www.astro.ucla.edu/~wright/CosmoCalc.html) adopting flat cosmology. (4) Isotropic H\textsubscript{2}O maser luminosity logarithm, $ \log_{10} (L_{\mathrm{H\textsubscript{2}O}} / L_{\sun}) $. (5) HCN(J=1-0) line luminosity logarithm, $\log_{10}$(L\textsubscript{HCN}/\textbf{K}kms\textsuperscript{-1}pc\textsuperscript{2}). (6) HCO+(J=1-0) line luminosity logarithm, $\log_{10}$(L\textsubscript{HCO+}/\textbf{K}kms\textsuperscript{-1}pc\textsuperscript{2}). (7) \textsuperscript{12}CO(J=1-0) line luminosity logarithm, $\log_{10}$(L\textsubscript{CO}/\textbf{K}kms\textsuperscript{-1}pc\textsuperscript{2}). (8) Normalized H\textsubscript{2}O maser emission luminosity (L\textsubscript{H\textsubscript{2}O}/L\textsubscript{CO}) logarithm. (9) Normalized HCN line luminosity (L\textsubscript{HCN}/L\textsubscript{CO}) logarithm. (10) Normalized HCO+ line luminosity (L\textsubscript{HCO+}/L\textsubscript{CO}) logarithm. (11) References for the H\textsubscript{2}O, HCN, HCO+ and CO luminosities, respectively, ** stands for not available observation or reference. 

{\bf References.} H\textsubscript{2}O data: (1) "https://safe.nrao.edu/wiki/bin/view/Main/PublicWaterMaserList", (2) \cite{Kuo2018}. HCN, HCO+ and CO data: (3) \cite{Liu2015}, (4) \cite{Knudsen2007}, (5) \cite{Lindberg2011}, (6) \cite{Baan2008}, (7) \cite{Sani2012}, (8) \cite{Imanishi2007}, (9) \cite{Imanishi2006}, (10) \cite{Koda2002}, (11) \cite{Israel2020}, (12) \cite{Costagliola2011}, (13) \cite{Green2016}, (14) \cite{Jiang2011}, (15) \cite{krause}, (16) \cite{McCoy2017}, (17) \cite{Querejeta}, (18) \cite{Imanishi2009}, (19) \cite{Curran2001}, (20) \cite{Raluy}, (21) \cite{Greve2009}.  

\end{tablenotes}
\end{tabular}

\end{table}
\end{landscape}
\twocolumn
 
 \begin{table}
 \caption{Correlation and regression results.}
     \begin{threeparttable}
    \begin{tabular}{lll}
  \hline\hline
  & \multicolumn{2}{l}{\textbf{For all data sample:}} \\
      & \underline{$\log_{10}$(L\textsubscript{HCN}/L\textsubscript{CO})}                    & \underline{$\log_{10}$(L\textsubscript{HCO+}/L\textsubscript{CO})} \\
$\log_{10}$(L\textsubscript{H\textsubscript{2}O}/L\textsubscript{CO})  & N = 30                          & N = 22  \\
& $\rho$ = 0.38    & $\rho$ = 0.66\\
& P = 1.9$\times$10\textsuperscript{-2} & P = 4.5$\times$ 10$^{-4}$\\
& S = 1.03    & S = 2.36    \\
& I = -6.67 & I = -4.77                   \\
& -6.67$\leq$ S $\leq$ 1.71  & 1.41 $\leq$ S $\leq$ 3.42\\
& d = 0.43& d = 0.53\\
\\* 
&  \multicolumn{2}{l}{\textbf{For sources with normal IR   luminosity:}} \\
& \underline{$\log_{10}$(L\textsubscript{HCN}/L\textsubscript{CO})}                    & \underline{$\log_{10}$(L\textsubscript{HCO+}/L\textsubscript{CO})}               \\
$\log_{10}$(L\textsubscript{H\textsubscript{2}O}/L\textsubscript{CO}) & N = 21                          & N = 15                      \\
              & $\rho$ = 0.56                        & $\rho$ = 0.83                    \\
              & P = 4.4$\times$ 10\textsuperscript{-3}                      & P= 5.8$\times$ 10\textsuperscript{-5}                  \\
              & S = 1.31                       & S = 2.71                 \\
              & I = -5.74                     & I = -3.53               \\
              & 0.22$\leq$ S $\leq$ 2.06            & 1.56 $\leq$ S $\leq$ 3.96       \\
              & d = 0.08                      & d = -0.04 \\
\hline\hline             
              \end{tabular}
\end{threeparttable}
\begin{tablenotes}
\item {\bf Note. N is the sample number. Spearman's rank correlation results: $\rho$ is the rank correlation coefficient between the variables and P is the P-value of the corresponding $\rho$, we consider the value P$\leq$0.05 for significant correlations. Kendall-Theil linear regression results: S is the slope of the regression line, I is the intercept, 95\% percent confidence interval of slope is given as a range of S, and the median deviation d.}
\end{tablenotes}
\end{table}

 \section{Discussion}
We expect H$_2$O maser emission to be correlated with the dense gas fraction in AGNs, for reasons introduced in Section~1. The dense gas fraction is usually expressed by the ratio of the line luminosity of one of the dense gas tracer molecules such as HCN, HCO+, CS, HNC, N\textsubscript{2}H+ to the line luminosity of one of the low-density gas tracers such as CO and C\,{\sc i} (neutral atomic carbon) lines. On the grounds that the CO emission line is the most effectively observed, we used it to calculate the dense gas fraction for our sample.

  As shown in Table \textsc{2}, both HCN and HCO+ dense gas fractions correlate with (L\textsubscript{H\textsubscript{2}O}/L\textsubscript{CO}) with Spearman's correlation coefficients equal to 0.38 and 0.66 respectively. However, HCO+ seems to be strongly correlated  with L\textsubscript{H\textsubscript{2}O}/L\textsubscript{CO} while L\textsubscript{HCN} / L\textsubscript{CO} shows a weaker correlation, with regression line equations
 
 \begin{equation} 
 \log_{10}(L\textsubscript{H\textsubscript{2}O} / L\textsubscript{CO}) = 1.03 \log_{10}(L\textsubscript{HCN} / L\textsubscript{CO}) - 6.67,
 \end{equation} 
 
 \begin{equation}
 \log_{10}( L\textsubscript{H\textsubscript{2}O} / L\textsubscript{CO}) = 2.36 \log_{10}(L\textsubscript{HCO+} / L\textsubscript{CO}) - 4.77.
 \end{equation}
 
  Fig. \textsc{1} and Fig. \textsc{2} display eq.(3) and eq.(4), respectively.
 \begin{figure}
\centering
 \vspace*{17pt}
\includegraphics[width=8cm]{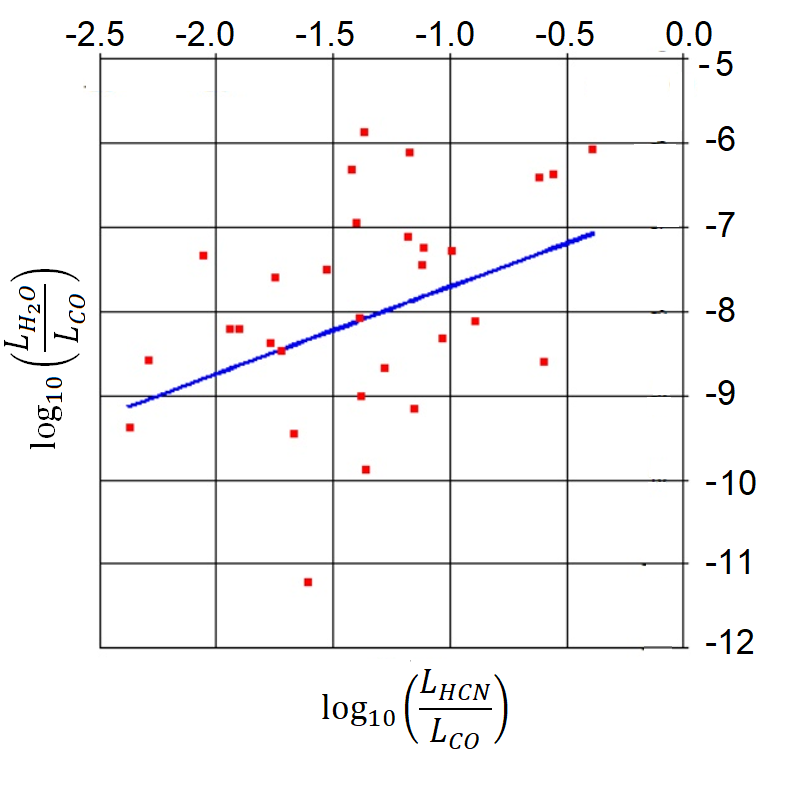}
\caption{  Relation between L\textsubscript{H\textsubscript{2}O}/L\textsubscript{CO} and L\textsubscript{HCN}/L\textsubscript{CO} in logarithm scale. The blue line is the Kendall-Their regression line of the relation (see eq.(3)).}
 \label{fig:fig1}
\end{figure} 

\begin{figure}
\centering
 \vspace*{17pt}
\includegraphics[width=8cm]{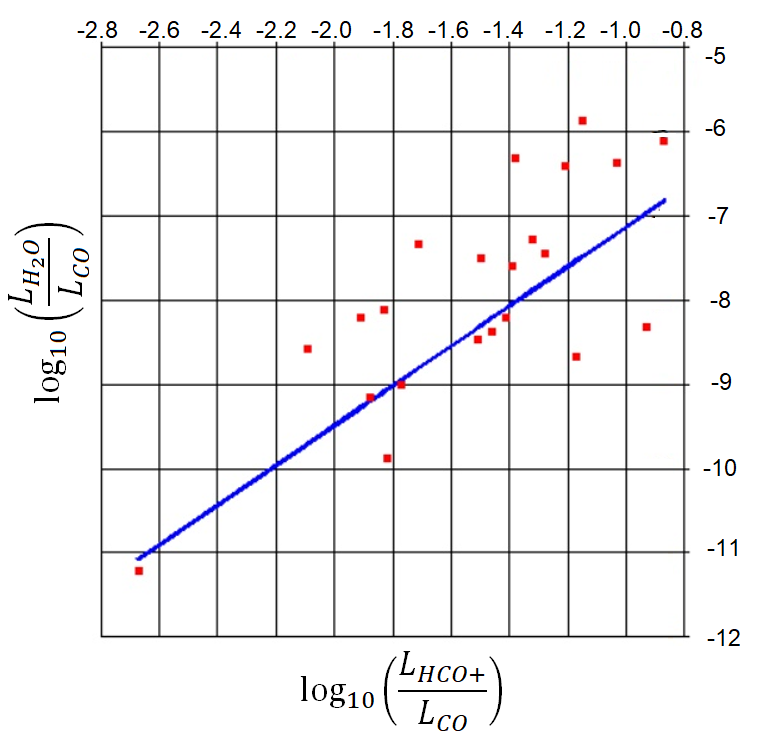}
\caption{Relation between L\textsubscript{H\textsubscript{2}O}/L\textsubscript{CO} and L\textsubscript{HCO+}/L\textsubscript{CO} in logarithm scale. The blue line is the Kendall-Theil regression line of the relation. A comparison with Fig.\textsc{1} clearly shows how HCO+ interrelates with maser emission stronger than HCN (see eq.(4)).}
 \label{fig:fig2}
\end{figure}

 We further investigated these correlations after eliminating the sources with only HCN measurements available. We find that there is no more correlation between L\textsubscript{H\textsubscript{2}O}/L\textsubscript{CO} and L\textsubscript{HCN}/L\textsubscript{CO} ($\rho$ = 0.27 and P = 0.23). This result casts doubts on HCN as an accurate and unbiased dense gas tracer in maser sources.
 
 Following the results reported by other authors (e.g. \citealt{Graci-Carpio-2006}, \citealt{Papadopoulos-2007}), the luminous and ultraluminous IR galaxies ((U)LIRG) show a noticeable enhancement of HCN that could not be necessarily related to dense gas, so it might not be a  useful tool to  use HCN as a dense gas tracer in these kind of galaxies. Rather, it is much safer to use HCO+ or high-level transitions of HCN and HCO+ such as HCN(J=2-1), HCN(J=3-2) and HCO+(J=3-2). We also tried to find the correlation after excluding (U)LIRG sources, which roughly represent only one third of the sample. L\textsubscript{HCN}/L\textsubscript{CO} was found to be moderately correlated with maser emission, with a sample number of 21 out of 30, and a $\rho$ = 0.56, P= 4$\times$10\textsuperscript{-3} as it is plotted in Fig. \textsc{3} and Fig. \textsc{4}. Also it is worth noticing that HCO+ correlates better with maser emission after removing (U)LIRG (N= 15, $\rho$ 0.83, and P= 6$\times$10\textsuperscript{-5}). This agrees with the results of  \cite{Papadopoulos-2007} and it may indicate the presence of turbulent molecular gas which can dilute the presence of HCO+ gas.\\
 Another factor that may affect the abundance of HCN relative to HCO+ is metallicity of the emitting source, in fact the abundance of HCN as a nitrogen bearing molecule decreases faster than HCO+ in low-metallicity CND environments (\citealt{refId0}, \citealt{Kepley2018}). Unfortunately, we cannot currently rule out this effect due to the small sample size but it would be possible to check for the metallicity effect on HCN/HCO+ ratios both at nuclei and at CND for H\textsubscript{2}OMM galaxies in further studies.

\begin{figure}
\centering
\vspace*{8pt}
\includegraphics[width=8cm]{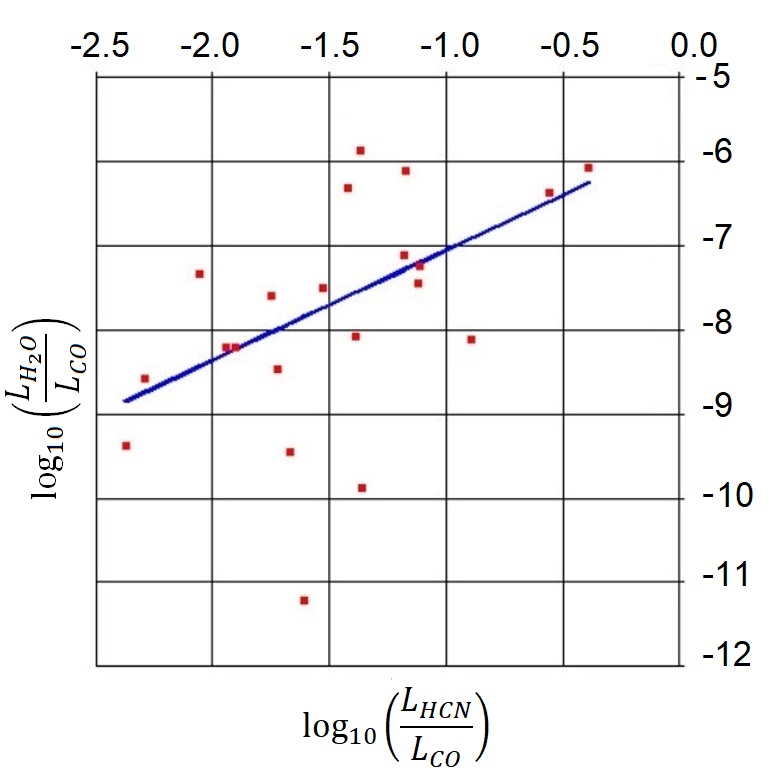}
\caption{Relation between maser emission and the dense gas fraction as extracted from HCN for normal IR radiation galaxies. The blue line is the Kendall$-$Theil regression line of the relation. (See Table \textsc{2})}
\label{fig:fig3}
\end{figure}

\begin{figure}
\centering
\vspace*{8pt}
\includegraphics[width=8cm]{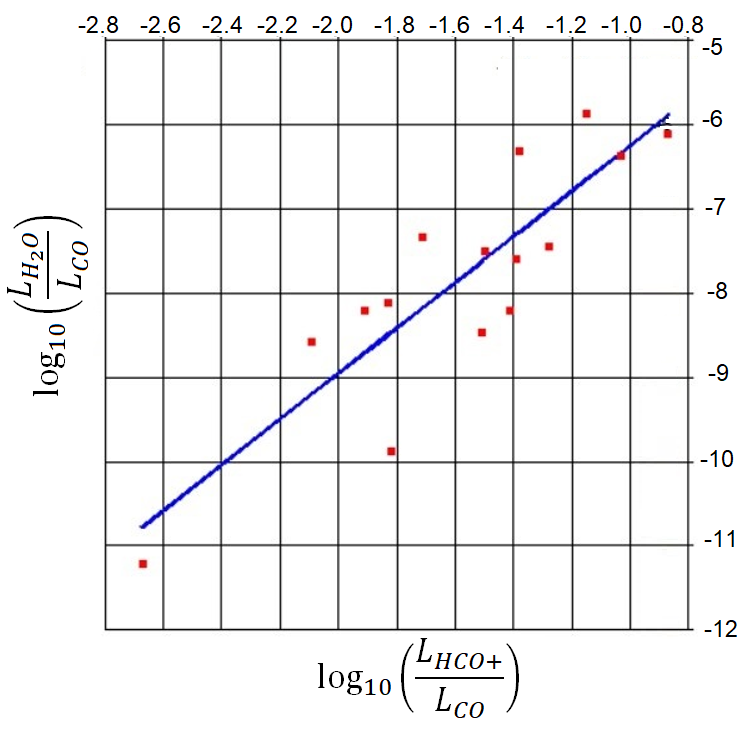}
\caption{Relation between maser emission and the dense gas fraction as extracted from HCO+ for normal IR radiation galaxies. The blue line is the Kendall$-$Theil regression line of the relation. (See Table \textsc{2})}
\label{fig:fig4}
\end{figure}

Fig. \textsc{5} and Fig. \textsc{6} show the box and whisker plot of maser types versus the dense gas luminosity fraction of HCN and HCO+, respectively. As it can be seen, the HCO+ dense gas fraction is slightly better distinguished between the two types of masers than the HCN dense gas fraction, with a tendency to be higher in MM sources in both cases.\\
In addition, a comparison of the ranges of the dense gas ratios in Fig. \textsc{5} and Fig. \textsc{6} immediately reveals the widest range is for HCN dense gas fraction in KM, this might be fairly interpreted as high enhancement of HCN in KMs. \cite{Graci-Carpio-2006} discussed observational evidences of HCN enhancement in high-mass star-formation regions, and since KMs are generally powered by starbursts, thus the embedded massive star-formation regions in galaxies with KM emission might be the reason of the wide range of HCN dense gas ratio.

Besides, the critical density of HCN was found to be two order of magnitude less than what is commonly accepted (n\textsubscript{crit} $\sim$ 10\textsuperscript{5}~cm\textsuperscript{-3}) (\citealt{kauffmann}). If this is the case, then HCN should not be considered as a dense gas tracer in extragalactic sources.
 
 \begin{figure}
\centering
 \vspace*{8pt}
 \includegraphics[width=8cm]{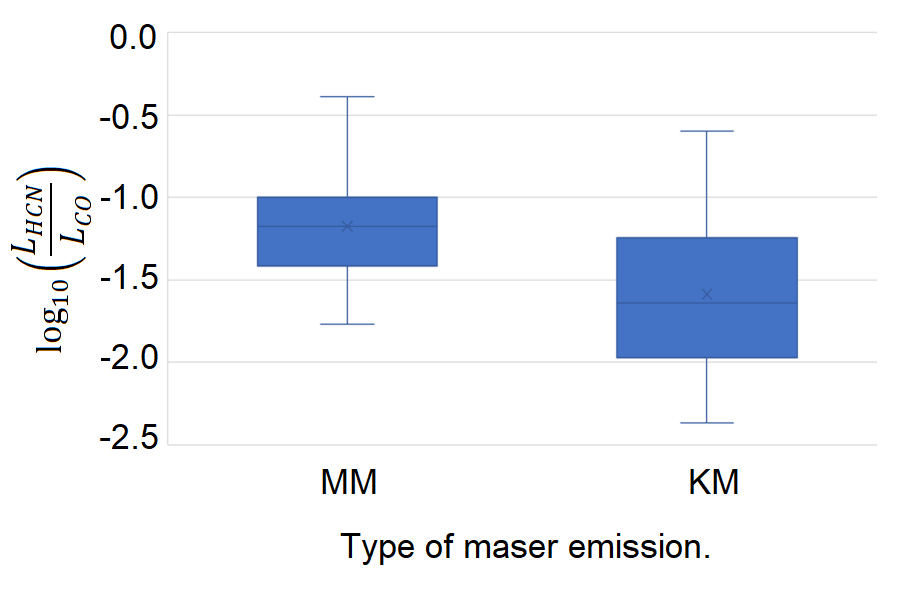}
 \caption{The box and whisker plot of maser types versus the HCN dense gas fraction in both type of megamaser galaxies: kilomasers (KM) and megamasers (MM).}
 \label{fig:fig5}
\end{figure}

\begin{figure}
\centering
 \vspace*{8pt}
 \includegraphics[width=8cm]{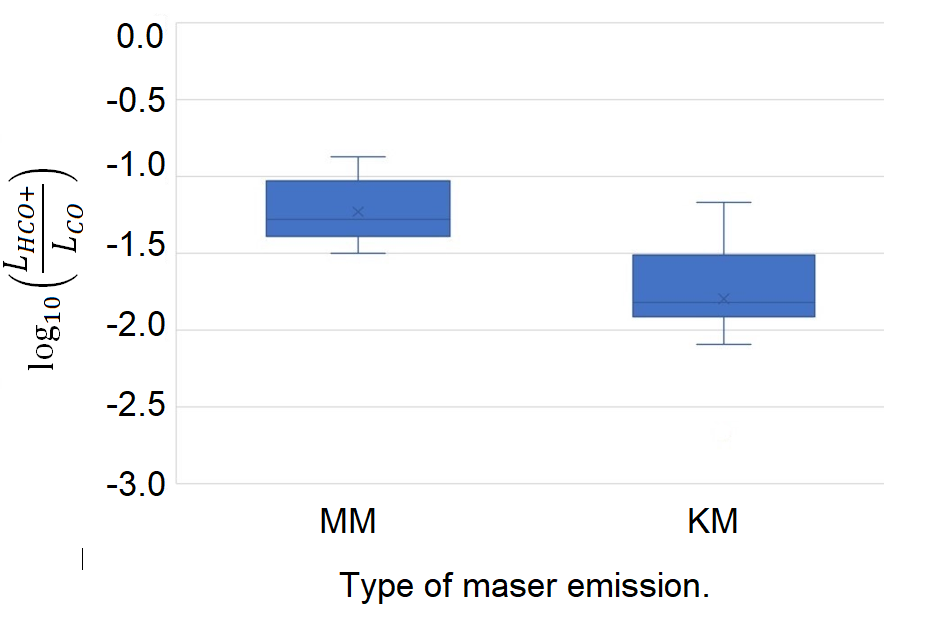}
 \caption{The box and whisker plot of maser types versus the HCO+ dense gas fraction in both type of megamaser galaxies: kilomasers (KM) and megamasers (MM). When compared with Fig. \textsc{5}, the tow types of maser are more distinguished here.}
 \label{fig:fig6}
\end{figure}

 Consequently, our results show that HCO+ rather than HCN is the molecular gas to be looked for if we want to place constraints on sources to survey for megamasers. 
 
\section {Conclusions}

HCN and HCO+ dense gas fractions in 30 water vapor megamaser galaxies have been statistically analysed. Both fractions show a correlation with water maser emission, with HCO+ being stronger than HCN. Moreover, after removing (U)LIRG sources, the HCO+ correlation becomes very tight and HCN correlation becomes stronger. Our results agree with previous works on dense gas as a favorable parameter to be used in water megamaser surveys performed by \cite{Zhang-2009} and \cite{HUANG}. According to our analyses, HCO+ is  preferable on HCN as a dense gas tracer in megamaser sources. A further study for high level transitions would help to better elucidate the circumstances in which each of these dense gas tracers could be reliable as a tracer to dense gas in maser vigorous sources.

\section*{Acknowledgements}
We thank Prof. Rafik Kandalyan for his valuable comments and discussions on the subject. E.N. Ercan would like to thank Bogazici University BAP for financial support under project code 18B03P1.

\section*{Data availability}
The data used in this study will be made available by the corresponding authors upon request.





\bsp	
\label{lastpage}
\end{document}